\documentclass[runningheads]{llncs}
\usepackage[T1]{fontenc}
\usepackage{graphicx}
\usepackage{graphicx}

\usepackage{amssymb}
\usepackage{amsmath}
\usepackage{mathtools}
\usepackage{bbm}
\usepackage{enumitem}
\usepackage{tabularx}
\usepackage{booktabs}
\usepackage{multirow}
\usepackage{amsfonts}
\usepackage{makecell}
\usepackage{bm}
\usepackage{cite}
\usepackage{array}
\usepackage{url}
\usepackage[misc]{ifsym}
\usepackage{bbding}

\usepackage[colorlinks,
linkcolor=blue,
anchorcolor=blue,
citecolor=blue]{hyperref}

\makeatletter
\def\thanks#1{\protected@xdef\@thanks{\@thanks\protect\footnotetext{#1}}}
\makeatother

\begin{document}
\title{Towards Multi-modality Fusion and Prototype-based Feature Refinement for Clinically Significant Prostate Cancer Classification in Transrectal Ultrasound}
\titlerunning{Multi-modality TRUS for PCa Classification}
%
\author{
Hong Wu\inst{1,\dagger}\thanks{$\dagger$ Hong Wu and Juan Fu contribute equally to this work.} \and 
Juan Fu\inst{2,\dagger} \and
Hongsheng Ye\inst{2} \and
Yuming Zhong\inst{1}\and
Xuebin Zou\inst{2}\and
Jianhua Zhou\inst{2,}\Envelope 
\and Yi Wang\inst{1,}\Envelope\thanks{\Envelope Corresponding authors: Jianhua Zhou and Yi Wang.}
}
\authorrunning{H. Wu et al.}
%
\institute{
Smart Medical Imaging, Learning and Engineering (SMILE) Lab,
Medical UltraSound Image Computing (MUSIC) Lab,
School of Biomedical Engineering,
Shenzhen University Medical School,
Shenzhen University, Shenzhen, China \and
The Department of Ultrasound, State Key Laboratory of Oncology in South China,
Guangdong Provincial Clinical Research Center for Cancer,
Sun Yat-sen University Cancer Center, Guangzhou, China\\
\email{zhoujh@sysucc.org.cn}, \email{onewang@szu.edu.cn}
}
\maketitle              
\begin{abstract}
Prostate cancer is a highly prevalent cancer and ranks as the second leading cause of cancer-related deaths in men globally.
Recently, the utilization of multi-modality transrectal ultrasound (TRUS) has gained significant traction as a valuable technique for guiding prostate biopsies.
In this study, we propose a novel learning framework for clinically significant prostate cancer (csPCa) classification using multi-modality TRUS.
The proposed framework employs two separate 3D ResNet-50 to extract distinctive features from B-mode and shear wave elastography (SWE).
Additionally, an attention module is incorporated to effectively refine B-mode features and aggregate the extracted features from both modalities.
Furthermore, we utilize few shot segmentation task to enhance the capacity of classification encoder.
Due to the limited availability of csPCa masks, a prototype correction module is employed to extract representative prototypes of csPCa.
The performance of the framework is assessed on a large-scale dataset consisting of 512 TRUS videos with biopsy-proved prostate cancer.
The results demonstrate the strong capability in accurately identifying csPCa, achieving an area under the curve (AUC) of 0.86.
Moreover, the framework generates visual class activation mapping (CAM), which can serve as valuable assistance for localizing csPCa.
These CAM images may offer valuable guidance during TRUS-guided targeted biopsies, enhancing the efficacy of the biopsy procedure.
\textit{The code is available at}
\url{https://github.com/2313595986/SmileCode}.

\end{abstract}
\section{Introduction}
Prostate cancer (PCa) is one of the most prevalent malignancies among men~\cite{siegel2023cancer}.
Prostate biopsy is the most reliable method for PCa diagnosis.
Based on the microscopic appearance of prostate tissue,
the results can be divided into benign prostatic hyperplasia (BPH),
clinically significant PCa (csPCa)
and clinically insignificant PCa (cisPCa)~\cite{matoso2019defining}.
Since csPCa generally exhibits a worse prognosis and requires timely treatment,
identifying csPCa patients helps improve the survival rate.

\begin{figure}[t]
\centering
\includegraphics[width=\textwidth]{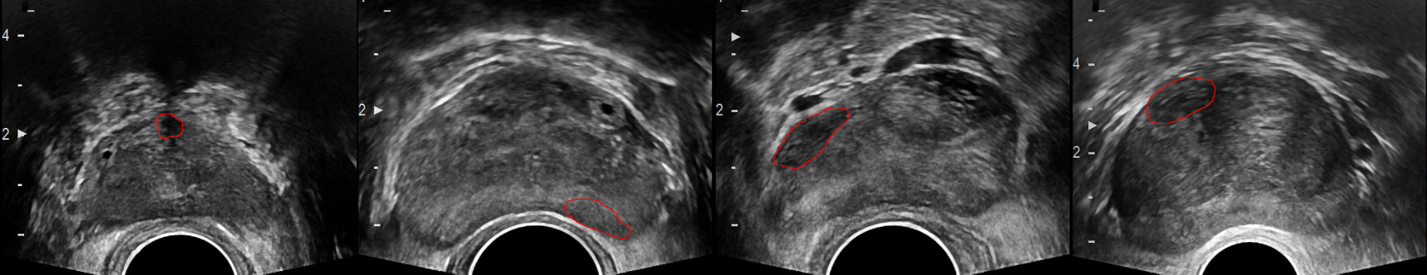}
\caption{
 Visualization of csPCa masks (red) on the specific frames from 4 TRUS videos. }
 \label{fig:csPCa_mask}
\end{figure}

According to the current guidelines, any suspected PCa patients should have a multiparametric magnetic resonance imaging (mp-MRI) before biopsy~\cite{albright2015prostate}.
Prostate mp-MRI is capable of depicting localization and morphology of PCa, thereby facilitating the implementation of the biopsy.
However, the widespread implementation of prostate mp-MRI is restricted due to complex operations and high cost.
Transrectal ultrasound (TRUS) serves as a commonly used modality for the guidance of prostate biopsies~\cite{grey2022multiparametric}, yet the low contrast TRUS poses challenges for clinicians to identify PCa regions (see Fig.~\ref{fig:csPCa_mask}).
It inevitably results in more unnecessary biopsies,
with a higher risk of complications such as rectal bleeding~\cite{loeb2013systematic}.
Hence, it is crucial to develop effective strategies for csPCa detection in TRUS. 

Previous study~\cite{ahmad2013transrectal} indicated that regions assigned a Gleason score of 7 exhibited statistically higher Young's modulus compared to regions with a Gleason score of 6.
Building upon this finding, 
Wildeboer~\textit{et al.}~\cite{wildeboer2020automated} developed a random forest classifier analyzing B-mode, SWE and contrast-enhanced ultrasound, reached area under curves (AUC) of 0.75 for PCa.
Liang~\textit{et al.}~\cite{liang2021nomogram} utilized a radiomics model that incorporated B-mode and SWE to classify the PCa and achieved an AUC of 0.85.
Notably, the above two methods required the biopsy pathology as reference to draw all region of interests (ROIs).
Recently, deep learning method has been applied to analyze TRUS videos for identifying csPCa.
Sun~\textit{et al.}~\cite{sun2023three} proposed a prostate mask guided hierarchical framework to identify csPCa,
and achieved an AUC of 0.85 in the external validation set.

In this study, we propose a multi-modality TRUS video classification network for the accurate identification of csPCa patients.
Inspired by the multi-task learning strategy,
a limited number of segmentation masks of csPCa are reconstructed based on biopsy pathology (see Fig.~\ref{fig:csPCa_mask}) and employed to train an auxiliary task for enhancing the encoder of classification network.
Experiments on a large-scale multi-modality TRUS video dataset demonstrate the efficacy of our method.
Our contributions can be summarized as follows:
\begin{itemize}
	\item[$\bullet$] To our knowledge, this is the first deep learning based framework for the classification of csPCa in multi-modality TRUS videos.
	More important, the framework generates visual CAM images, which may offer valuable guidance for TRUS-guided targeted biopsies.
	\item[$\bullet$] To better leverage the multi-modality information, we propose an attention module to optimize B-mode features, and fuse the features of B-mode and SWE.
	\item[$\bullet$] We add segmentation task as an auxiliary task to enhance the classification encoder, and employ a prototype correction module to address the issue of limited csPCa segmentation masks.
\end{itemize}

\section{Method}

\begin{figure}[t]
\centering
\includegraphics[width=\textwidth]{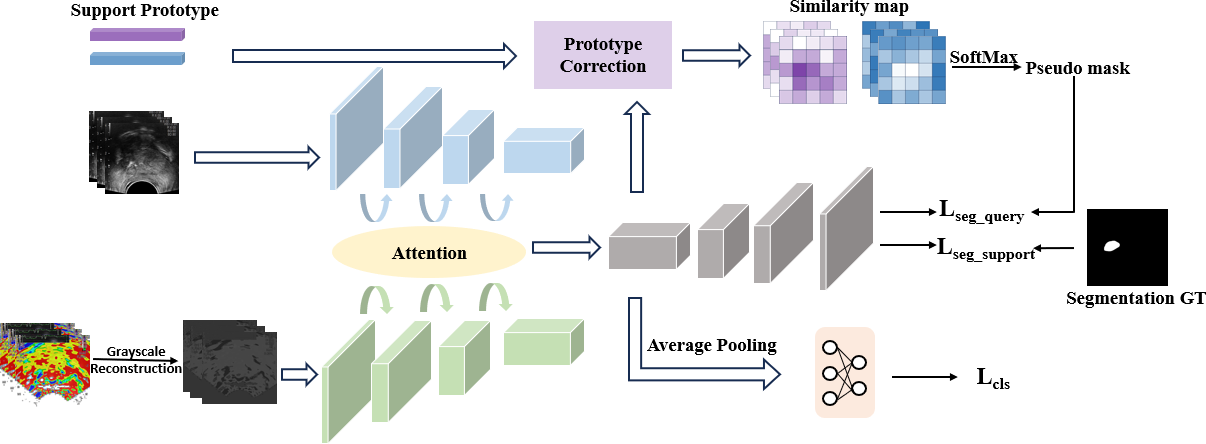}
\caption{Schematic overview of our method for identification of csPCa.
Segmentation serves as an auxiliary task for enhancing classification performance.}
\label{fig:workflow}
\end{figure}

The proposed framework is illustrated in Fig.~\ref{fig:workflow}.
Two 3D ResNet-50~\cite{he2016deep} are employed to extract features from B-mode and SWE respectively and perform the classification task.
We employ a dimensional attention module to refine the extracted features from B-mode and adopt an adaptive spatial attention module to fuse the features from two modalities (Fig.~\ref{fig:module} left).
Furthermore, we add a 3D U-Net decoder~\cite{cciccek20163d} with skip-connection to perform the segmentation task.
The segmentation task serves as an auxiliary task to enhance the encoder.
Given the limited availability of csPCa masks, we train the segmentation network in a few-shot paradigm (Fig.~\ref{fig:module} right).

\subsection{Dimensional Attention Module}
To enforce the network's attention towards the csPCa regions,
we introduce a dimensional attention module~\cite{wu2019weakly},
as shown in the left of Fig.~\ref{fig:module}.
It utilizes dimension-wise attention to re-weight the features and fuses the re-weighted features afterwards.
Specifically,
$F_X \in \mathbb{R}^{H \times W \times T \times C}$
can be viewed as $C$ feature cubes.
For each feature cube, we first apply average pooling to its slices along the three different dimensions then obtain three vectors
$F^H \in \mathbb{R}^{H}$, 
$F^W \in \mathbb{R}^{W}$
and $F^T \in \mathbb{R}^{T}$.
Subsequently, squeeze-and-excitation~\cite{hu2018squeeze} is employed to compute the attention scores:
\begin{equation}
\label{eq:attention}
s_D = \sigma(FC(F^D)),
D \in \{H, W, T\},
\end{equation}
where $FC$ is fully connected layer and $\sigma$ is sigmoid operation.
Then we re-weight the features by the attention scores and fuses them together.

\subsection{Adaptive Spatial Attention Module}
Abnormal hypoechogenicity in B-mode~\cite{schoots2015magnetic} and 
abnormal stiff regions in SWE~\cite{ahmad2013transrectal} can be considered as an indicator of csPCa.
Motivated by the recent advancements in attention mechanisms for feature aggregation~\cite{huang2023joint, wang2023a2fseg},
we introduce an adaptive spatial attention module to aggregate the features from two modalities in pixel-level.

\begin{figure}[t]
	\centering
	\includegraphics[width=\textwidth]{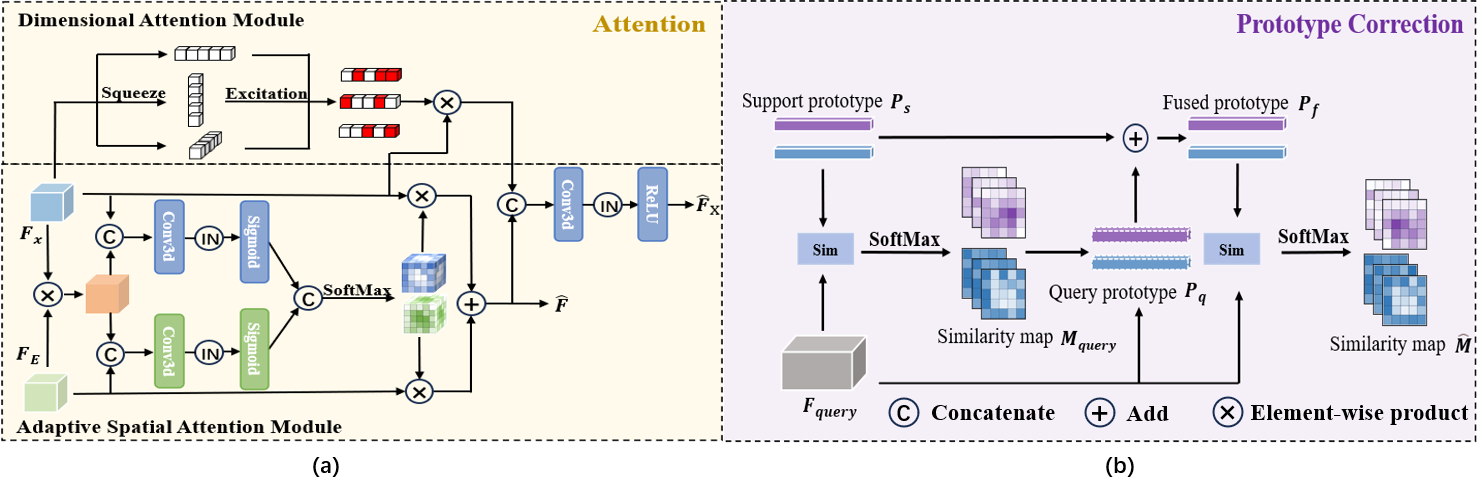}
	\caption{Architecture of the attention module and prototype correction module.}
	\label{fig:module}
\end{figure}

As shown in Fig.~\ref{fig:module} (a),
we first compute the element-wise product of the features $F_X$ (B-mode) and $F_E$ (SWE).
Then the element-wise product are concatenated with the modality-specific features and pass through the convolutional blocks to provide modality-specific attention weights.
Subsequently, we multiply the attention weights with the corresponding modality feature maps to generate the fused feature maps $\hat{F}$.
As depicted in Fig.~\ref{fig:workflow}, 
$\hat{F}$ are subsequently integrated back into the SWE branches.
For the B-mode branches, $\hat{F}$ are concatenated with dimensional attention module output, and then pass through the convolution block before integration.

\subsection{Few Shot Segmentation for csPCa}
Considering the interplay between the tumor classification and segmentation tasks~\cite{amyar2020multi, zhou2021multi}, 
we add the segmentation task as an auxiliary task to assist the classification network to extract more discriminative features.
Due to the limited availability of csPCa masks,
we also introduce a prototype correction module
to extract representative prototypes of csPCa.

As shown in Fig.~\ref{fig:workflow}, we add a 3D U-Net decoder 
with skip-connection to perform the segmentation task.
Specifically, 
we denote the csPCa set with mask as support set $(X_{support}, Y_{support})$ and the csPCa set without mask as query set $X_{query}$.
For the non-csPCa set, we denote it as $X_{neg}$. 

For the classification task,
the loss function is defined as:
\begin{equation}
\label{eq:loss_cls}
\mathcal{L}_{cls} = - Y_{cls} \log(\hat{y}_{cls}) - (1 - Y_{cls}) \log(1 - \hat{y}_{cls}),
\end{equation}
where $\hat{y}_{cls}$ is the classification prediction of the network and $Y_{cls}$ represents the csPCa classification label.

For the support set in the segmentation task,
we minimize the cross entropy loss:
\begin{equation}
\label{eq:fusion}
\mathcal{L}_{\text{seg\_support}} = - Y_{support} \log(\hat{y}_{seg}) - (1 - Y_{support}) \log(1 - \hat{y}_{seg}),
\end{equation}
where $\hat{y}_{seg}$ denotes predicted probability map for $X_{support}$.
Since the non-csPCa set does not contain csPCa regions, 
we employ zero supervision for its segmentation:
\begin{equation}
\label{eq:fusion1}
\mathcal{L}_{\text{seg\_neg}} = - log(1 - \hat{y}_{seg}),
\end{equation}
For the query set, we adopt the few shot segmentation paradigm~\cite{dong2018few, liu2020prototype} to generate pseudo label and use them to supervise segmentation network training.
Concretely,
we perform average pooling operation within the csPCa mask $Y_{support}$ and
compute support prototype $P_{S}$ from the features of $X_{support}$:
\begin{equation}
\begin{split}
\label{eq:prototype1}
P_{S} &= \frac{1}{K}\sum_{k=1}^{K}MAP(f_{\theta}(X_{support}^k), Y_{support}^{k}) \\&= 
\frac{1}{K}\sum_{k=1}^{K}\frac{\sum_{}F_{support}^{k} \cdot Y_{support}^{k}}
{\sum_{}Y_{support}^{k}},
\end{split}
\end{equation}
where $K$ denotes the size of $X_{support}$,
$f_{\theta}$ denotes the encoder
and $F_{support}^{k}$ is the extracted support feature.
Then we generate the cosine similarity maps between extracted query feature and support prototype:
\begin{equation}
\label{eq:prototype2}
M_{query} = \sigma(sim({f_{\theta}(X_{query}), P_{S}})),
\end{equation}
where $\sigma$ denotes softmax operation and $sim(\cdot, \cdot)$ is the cosine similarity function.

However, the support prototype $P_{S}$ are biased to represent the csPCa class duo to the limited number of csPCa masks.
We introduce a prototype correction module to refine the support prototype without introducing extra training parameters.
As shown in Fig.~\ref{fig:module} (b), 
we generate the query prototype based on the similarity map $M_{query}$ and the query features,
then fuse it with the support prototype:
\begin{equation}
\label{eq:prototype}
P_f = \frac{1}{2}(MAP(F_{query}, M_{query}) + P_s).
\end{equation}
Finally, we rank the similarity and select the top 100 points as pseudo label,
with half chosen from the foreground similarity map and the other half from the background similarity map.
Then minimize the cross entropy at pixel-level:
\begin{equation}
\mathcal{L}_{seg\_query} = 
-\frac{1}{|\mathcal{P}|}
\sum_{\mathclap{\vec{j_p} \in \mathcal{P}}}
log(\hat{Y}_{seg}(\vec{j_p}))
-\frac{1}{|\mathcal{N}|}
\sum_{\mathclap{\vec{j_n} \in \mathcal{N}}}
log(1-\hat{Y}_{seg}(\vec{j_n}))
,
\end{equation}
where $\mathcal{P}$ denotes the set of points sampled as foreground, 
$\mathcal{N}$ denotes the set of points sampled as background.

In summary, 
we employ the sum of the classification loss $\mathcal{L}_{cls}$ and
the segmentation loss $\mathcal{L}_{seg}$
to train the network:

\begin{equation}
\label{eq:loss_total}
\mathcal{L} = \mathcal{L}_{cls} + \mathcal{L}_{seg\_support} + \alpha\cdot\mathcal{L}_{seg\_query} + \beta\cdot\mathcal{L}_{seg\_neg},
\end{equation}
where $\alpha$ and $\beta$ are weighting coefficients.

\begin{table}[t]
	\centering                                     
	\caption{Comparison of different methods (best results are highlighted in bold).}
	\label{tab:result}
	\small
	\begin{tabular}{cccccccccc}
		\toprule
		B-mode     & SWE        & Attention     & Multi-task    &PCM        & AUC      & F1       & Acc &Sen  &Spe\\
		\midrule
		\checkmark &            &            &          &                        & 0.74     & 0.81     & 0.68  &\textbf{0.99}   &0.08\\
		       &\checkmark   &            &          &                        & 0.59     & 0.80     & 0.68  &0.97   &0.11\\
		\checkmark & \checkmark &            &          &                        & 0.79     & 0.81     & 0.71  &0.87   &0.41\\
		\checkmark & \checkmark & \checkmark &          &                        & 0.83    &  0.86   &0.79    &  0.93 &  0.51 \\ 
		\checkmark & \checkmark & \checkmark & \checkmark  &                     &0.84      & 0.86     &0.81   &0.87   &0.68\\
        \checkmark & \checkmark & \checkmark & \checkmark  & \checkmark           
		& \textbf{0.86}     & \textbf{0.87}         & \textbf{0.81}  & 0.86
 & \textbf{0.73}\\   
		\midrule
		\multicolumn{5}{c}{Sun~\textit{et al.}~\cite{sun2023three} w/o prostate mask}  
		& 0.78     & 0.80     & 0.69  &0.55  &0.95\\
		\bottomrule
	\end{tabular}
\end{table}

\section{Experiments}

\subsubsection{Dataset.}
We performed experiments on an in-house multi-modality TRUS video dataset collected from the Cancer Center of Sun Yat-Sen University.
This study included a cohort of 512 patients who underwent B-mode ultrasound, SWE, and prostate biopsy procedures.
Among them, 346 patients were diagnosed with csPCa,
while 166 presented with non-csPCa (BPH or cisPCa).
We randomly employed 404 scans (275 with csPCa) for training,
and 108 scans (71 with csPCa) for testing.
The TRUS videos were resized to $200 \times 144 \times 144$ for reducing computational cost.
The intensities of B-mode within each video were normalized to [0, 1].
The SWE images were underwent grayscale reconstruction according to~\cite{xiao2014computer}.

The segmentation annotation of csPCa is very difficult to obtain.
Firstly, pathologists examined specimens under a microscope and identified the spatial location of the csPCa on TRUS.
Subsequently, ultrasound radiologists used it as a reference to annotate the locations of cancerous lesions.
Segmentation masks of csPCa in 4 scans were manually annotated by ultrasound radiologists using ITK-SNAP.
These 4 labeled samples in training set were used as support set $(X_{support}, Y_{support})$,
and 271 unlabeled positive samples were query set $X_{query}$.

\subsubsection{Implementation Details.}
The method was implemented on PyTorch, using an NVIDIA RTX8000 GPU with 48G memory.
The network was optimized by a stochastic gradient descent optimizer for 200 epochs, with an initial learning rate of 0.0001.
The ploy learning policy was used, 
$(1 - \text{epoch} / 200)^{0.9}$.
To address the class imbalance issue, the batch size was set to 2, with each batch consisting of a csPCa and a non-csPCa.
For loss function, $\alpha$ and $\beta$ were set to 0.001 and 0.1, respectively.

\begin{figure*}[t]
	\centering
	\includegraphics[width=0.8\textwidth]{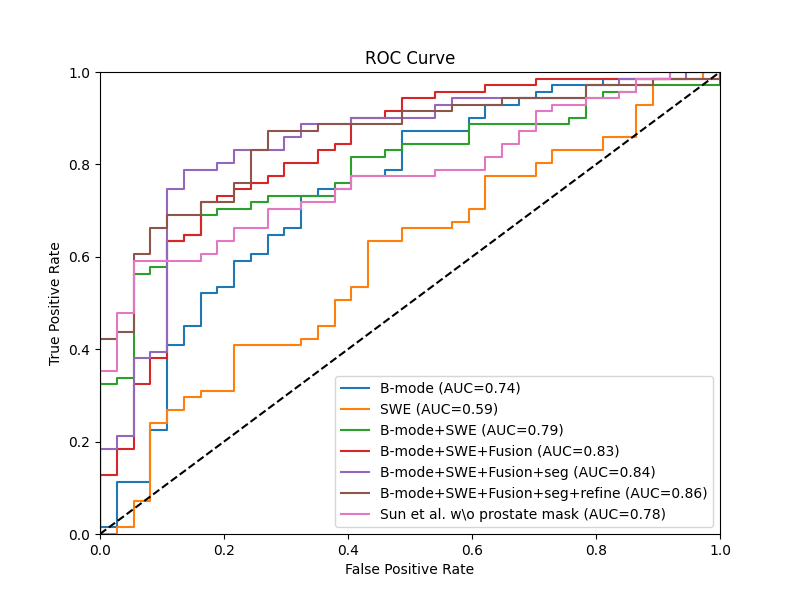}
	\caption{Comparison of ROC curves for different methods on the testing set.}
	\label{fig:roc}
\end{figure*}

\subsubsection{Quantitative and Qualitative Results.}
To evaluate the classification performance of our framework, we employed five evaluation metrics: area under the ROC curve (AUC), F1-score (F1), accuracy (Acc), sensitivity (Sen), and specificity (Spe).
Higher scores of these metrics show better performance.

To evaluate the impact of each component in our method, we conducted ablation analyses on single modality, where we used only B-mode or SWE information.
Additionally, we performed ablation experiments to assess the influence of the attention module, auxiliary task, and prototype correction module (PCM) on the overall performance of our approach.
These ablation studies allowed us to determine the contribution of each component in improving the classification outcomes.
Quantitative results are listed in Table~\ref{tab:result} and the ROC curves are depicted in Fig.~\ref{fig:roc}.
Our method achieved an AUC of 0.86, F1-score of 0.87, accuracy of 0.81, sensitivity of 0.86 and specificity of 0.73, demonstrating its strong performance in csPCa classification.
It can be observed that concatenating the two modalities (B-mode and SWE) resulted in a notable improvement in AUC compared to using either modality individually.
This proves the importance of leveraging the complementary information provided by multiple modalities to improve the classification performance.
Moreover, the addition of the fusion module improved the accuracy of the model,
and the addition of the segmentation task and the PCM has largely improved the specificity of the model.

We also compared our method with the most relevant work by~\cite{sun2023three}.
Our method outperformed theirs in terms of AUC by 0.08 when the prostate segmentation module was excluded, 
demonstrating its superiority in classifying samples without relying on prostate segmentation masks.

\begin{figure*}[t]
	\centering
	\includegraphics[width=\linewidth]{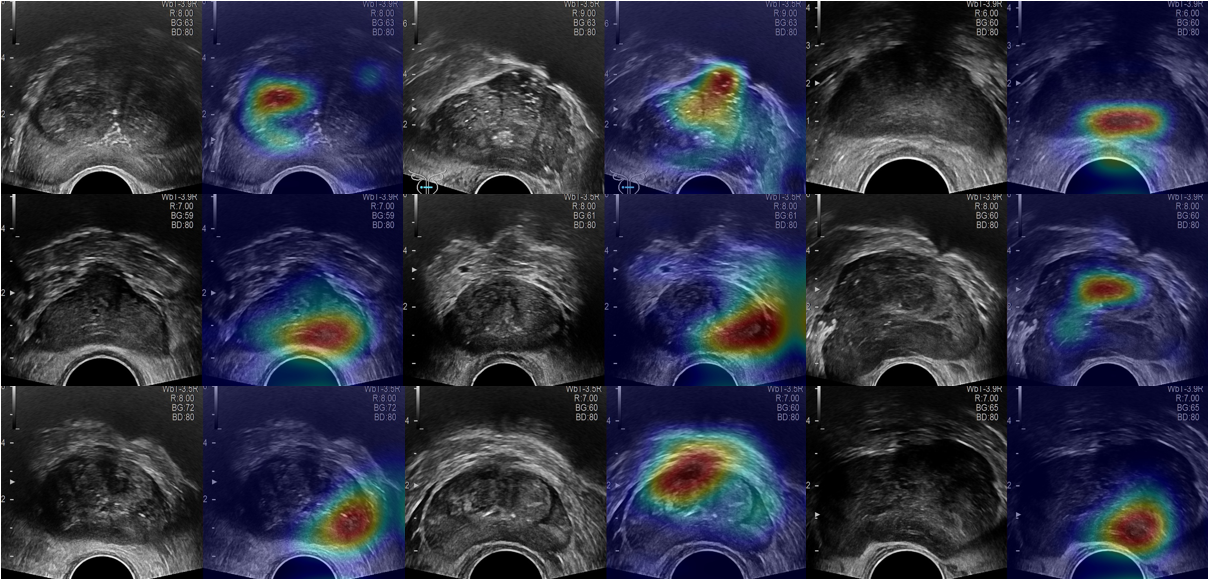}
	\caption{Nine examples illustrating TRUS frames and their corresponding CAM generated by our network. The ultrasound radiologists verified that the model's predictions of suspicious tumor regions were in agreement with the spatial locations provided by the pathologists.}
	\label{fig:cam}
\end{figure*}

To identify the frames and specific locations within TRUS videos that played a significant role in network's prediction of csPCa, we utilized gradient-weighted class activation mapping (Grad-CAM)~\cite{selvaraju2017grad}.
This approach allowed us to generate heat maps, visualizing the regions of interest and their contribution to the prediction, as shown in Fig.~\ref{fig:cam}.
Ultrasound radiologists confirmed that the suspicious tumor regions predicted by the model were consistent with the spatial location offered by pathologists.
This finding suggests the potential feasibility of TRUS-guided targeted biopsy, by leveraging the CAM images generated using our method.

\section{Conclusion}
We introduce a multi-task learning framework for the recognition of csPCa in multi-modality TRUS.
The primary attribute is to fully exploit B-Mode and SWE image information through the attention module and use few shot segmentation to help the encoder learn more representative features.
Experimental results demonstrate our proposed framework is effective in classifying and positioning csPCa.
This may provide useful information to assist doctors conducing TRUS-guided targeted biopsy.

%
%

\begin{credits}
\subsubsection{\ackname}
This work was supported in part by the National Natural Science Foundation of China under Grants 81971631, 82320108011 and 62071305,
in part by the Guangdong-Hong Kong Joint Funding for Technology and Innovation under Grant 2023A0505010021,
and in part by the Guangdong Basic and Applied Basic Research Foundation under Grant 2022A1515011241.
\end{credits}
%
%
%
%
\bibliographystyle{splncs04}
\bibliography{reference}
\end{document}